\documentclass[%
 reprint,
superscriptaddress,
 amsmath,amssymb,
 aps,
prapplied, floatfix, longbibliography ]{revtex4-2}
\usepackage{graphicx}
\usepackage{dcolumn}
\usepackage{bm}
\newcommand{\modR}[1]{                 {#1}}
\newcommand{\modB}[1]{                 {#1}}
\usepackage[colorlinks,linkcolor=blue,anchorcolor=blue,citecolor=blue,urlcolor=blue]{hyperref}
\begin{document}

\title{Unusual electric polarization behavior in
       elemental quasi-2D allotropes of selenium}

\author{Dan~Liu}
\email      {liudan2@seu.edu.cn}%
\affiliation{School of Physics,
             Southeast University,
             Nanjing, 211189, PRC}

\author{Lin~Han}
\affiliation{School of Physics,
             Southeast University,
             Nanjing, 211189, PRC}

\author{Ran~Wei}
\affiliation{School of Physics,
             Southeast University,
             Nanjing, 211189, PRC}

\author{Shixin~Song}
\affiliation{School of Physics,
             Southeast University,
             Nanjing, 211189, PRC}

\author{Jie~Guan}
\affiliation{School of Physics,
             Southeast University,
             Nanjing, 211189, PRC}

\author{Shuai~Dong}
\affiliation{School of Physics,
             Southeast University,
             Nanjing, 211189, PRC}

\author{David Tom\'{a}nek}
\email      {tomanek@msu.edu}%
\affiliation{Physics and Astronomy Department,
             Michigan State University,
             East Lansing, Michigan 48824, USA}

\date{\today}

\begin{abstract}
We investigate tunable electric polarization and electronic
structure of quasi-two-dimensional (quasi-2D) allotropes of
selenium, which are formed from their constituent one-dimensional
(1D) structures through an inter-chain interaction facilitated by
the multi-valence nature of Se. Our {\em ab initio} calculations
reveal that different quasi-2D Se allotropes display different
types of electric polarization, including ferroelectric (FE)
polarization normal to the chain direction in $\alpha$ and
$\delta$ allotropes, non-collinear ferrielectric (FiE)
polarization along the chain axis in $\tau$-Se, and
anti-ferroelectric (AFE) polarization in $\eta$-Se. The magnitude
and direction of the polarization can be changed by a previously
unexplored rotation of the constituent chains. In that case, an
in-plane polarization direction may change to out-of-plane in
$\alpha$-Se and $\delta$-Se, flip its direction, and even
disappear in $\tau$-Se. Also, the band gap may be reduced and
changed from indirect to direct by rotating the constituent chains
about their axes in these quasi-2D Se allotropes.
\end{abstract}




\maketitle
\renewcommand\thesubsection{\arabic{subsection}}




\section{Introduction}

\begin{figure*}[t]
\includegraphics[width=1.4\columnwidth]{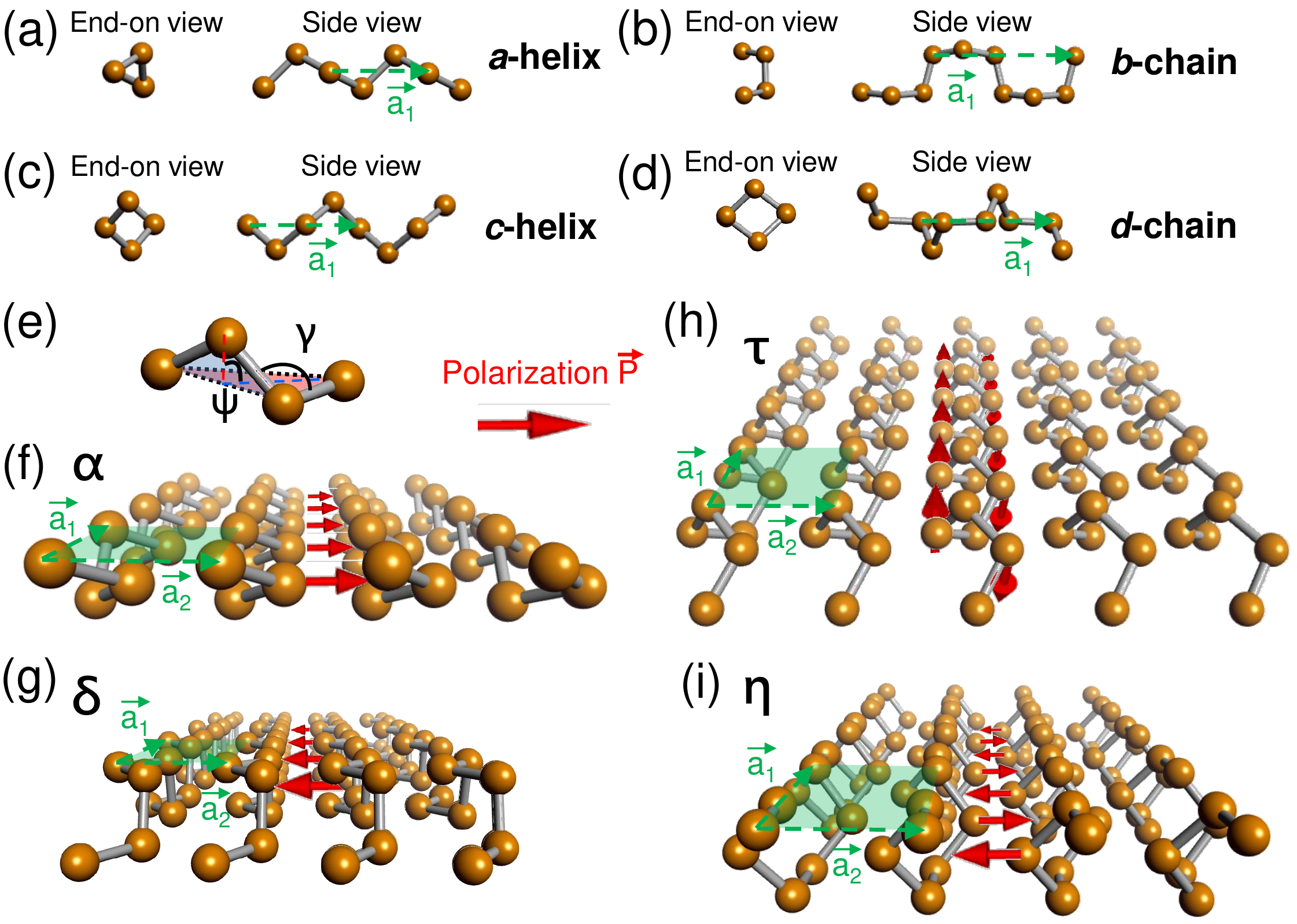}
\caption{%
Atomic structure of 1D allotropes of Se, namely %
(a) $a$-helix, %
(b) $b$-chain, %
(c) $c$-helix, and %
(d) $d$-chain. %
(e) Definition of the bond angle $\gamma$ and the dihedral angle
$\psi$ used to characterize the structures. Also shown is the
atomic structure of the corresponding quasi-2D allotropes %
(f) $\alpha$-Se, %
(g) $\delta$-Se, %
(h) $\tau$-Se, and %
(i) $\eta$-Se. %
Red arrows indicate the local electric polarization. The unit
cells of the quasi-2D structures are highlighted by the
transparent green areas in (f), (g), (h) and (i). %
\label{fig1}}
\end{figure*}

It is not widely known that the term {\em ferroelectric}
\modR{(FE)} has been coined already in 1912 by Erwin
Schr\"{o}dinger as an electric counterpart to a
ferromagnet~\cite{Schrodinger1912}. At low enough temperatures,
ferromagnetic or ferroelectric behavior is manifested by a
permanent magnetic or electric polarization of a material that
changes its direction upon applying an external field during a
hysteretic process. Less than a decade after Schr\"{o}dinger
dismissed -- immediately after proposing it -- the concept of
ferroelectricity as unrealistic, ferroelectric behavior \modR{was}
identified experimentally in Rochelle
salt~\cite{{Valasek1920},{Valasek1921}}.
There is a long way from Rochelle salt to new, low-dimensional
ferroelectrics that have captured the attention of the scientific
community by their complex behavior and their potential
application in electronics. This is particularly true for recently
discovered 2D `van der Waals' (vdW) layered structures with
ferroelectric behavior and different origins of
polarization~\cite{{Duan2020},{Zhu2017},{Liu2016},{Xiao2018},{Wu2017}}.
Best-known displacive ferroelectrics derive their behavior %
\modB{ from a relative displacement of sublattices, which is
characterized by the position of a representative
atom within the unit cell. }%
Changes in polarization, typically along a given direction, may be
caused by applying an external electric field,
such as in thin-film BaTiO$_{3}$~\cite{{Eom2004},{Duan2015}}. %
\modR{Only in specific 3D materials including
PbTiO$_3$~\cite{Vanderbilt2009}, a strain gradient has been shown
to modify the electrical polarization by an effect
sometimes dubbed `flexoelectricity'~\cite{Noheda2011}. } %
Flip of the polarization direction has been observed in 2D vdW
systems, such as ZrI$_{2}$, where shear strain has changed the
symmetry of the system~\cite{{Dong2021},{Li2021}}. %
\modR{In these systems, whether due to an external electric field
or mechanical strain, polarization changes have been limited to
changes in the magnitude, whereas the polarization direction
remained the same or flipped by $180^\circ$.} Capability to modify
both magnitude and direction of polarization will open a whole new
perspective for the application of \modR{2D FE} materials.



In this study, we investigate the polarization and band structure
of four quasi-2D allotropes of Se, which we call $\alpha$-Se,
$\delta$-Se, $\eta$-Se and $\tau$-Se. These quasi-2D allotropes
are formed from their constituent one-dimensional (1D) compounds
we call the $a$-helix, $b$-chain, $c$-helix and $d$-chain, which
possess a rotational degree of freedom about their axis. All atoms
in the isolated 1D structures have two neighbors at the same
distance and are structurally as well as electronically
equivalent. In the quasi-2D allotropes, however, differences arise
among individual atoms due to the inter-chain interaction. Our
{\em ab initio} density functional theory (DFT) calculations
reveal that the Bader charge of inequivalent Se sites deviates
from six valence electrons, indicating the possibility of a
macroscopic electric polarization.

As we will discuss further on, we find a FE polarization in
$\alpha$-Se and $\delta$-Se, whereas $\eta$-Se behaves as an
anti-ferroelecric (AFE). $\tau$-Se behaves as a non-collinear
ferrielectric (FiE), with FE polarization along the chain axis and
AFE polarization in the out-of-plane direction. Rotating each
constituent 1D chain of these quasi-2D Se allotropes changes the
atomic symmetry, causing a change in the electrical properties. As
for $\alpha$-Se and $\delta$-Se, the in-plane polarization can be
changed to the out-of-plane direction. At specific rotation angles,
$\tau$-Se distorts to a highly symmetric structure and its FiE
polarization disappears. Rotation of the constituent 1D chains in
the quasi-2D allotropes also changes the band structure. We observe
a modification of the fundamental band gap and a change from an
indirect to a direct band gap in $\delta$-Se.

\section{Computational Techniques}

Our calculations of the stability, equilibrium structure, and
energy changes during structural transformations have been
performed using the density functional theory (DFT) as implemented
in the {\textsc{VASP}}~\cite{VASP,VASPPAW} code. We used the
Perdew-Burke-Ernzerhof (PBE)~\cite{PBE} exchange-correlation
functional, augmented by the vdW correction using the DFT-D2
approach~\cite{DET-D2} to describe the inter-chain interaction.
Periodic boundary conditions have been used throughout the study,
with monolayers represented by a periodic array of slabs separated
by a 30~{\AA} thick vacuum region. The calculations were performed
using the projector augmented wave (PAW) method~\cite{VASPPAW} and
$500$~eV as energy cutoff. The reciprocal space has been sampled
by a fine grid~\cite{Monkhorst-Pack76} of $9{\times}11$~$k$-points
in the 2D Brillouin zone (BZ) of $\alpha$-Se,
$8{\times}8$~$k$-points in $\tau$-Se, and $5{\times}11$~$k$-points
in the $\delta$-Se and $\eta$-Se structures. All geometries have
been optimized using the conjugate gradient (CG)
method~\cite{CGmethod}, until none of the residual
Hellmann-Feynman forces exceeded $10^{-2}$~eV/{\AA}. \modR{The
polarization was calculated using the standard Berry phase
approach~\cite{{Vanderbilt93},{Resta94}} as implemented in the
{\textsc{VASP}} code.}

\begin{table*}[t!]
\caption{%
\label{table1}%
Results of our DFT-PBE-D2 calculations for the optimum bond angle
$\gamma$, the dihedral angle $\psi$, and the cohesive energy
$E_{coh}$ of the various Se allotropes investigated in our study. %
}%
\centering %
\scalebox{1.0} %
{\begin{tabular}{lcccccccc} %
\hline \hline
   \textrm{} %
 & \textrm{\textit{a}-helix} %
 & \textrm{\textit{b}-chain} %
 & \textrm{\textit{c}-helix} %
 & \textrm{\textit{d}-chain} %
 & \textrm{$\alpha$-Se}
 & \textrm{$\delta$-Se}
 & \textrm{$\tau$-Se}
 & \textrm{$\eta$-Se} \\
\hline%
  {$\gamma$ \rm{(degrees)}} %
& {103} %
& {104; 107} %
& {113} %
& {103; 106} %
& {101; 104} %
& {103; 105} %
& {109} %
& {106; 105} \\
\hline%
  {$\psi$ \rm{(degrees)}} %
& {100} %
& {101; 91} %
& {64} %
& {79; 100} %
& {99; 103} %
& {98; 103} %
& {43; 78} %
& {77; 99} \\
\hline%
  {$E_{coh}$ \rm{(eV/atom)}} %
& {3.359} %
& {3.376} %
& {3.361} %
& {3.373} %
& {3.476} %
& {3.488} %
& {3.462} %
& {3.498} \\
\hline \hline
\end{tabular}}
\label{table1}
\end{table*}

\section{Results}

\subsection{Atomic structure of quasi-2D Se allotropes and their
             electric polari\-zation}


One characteristic of group-VI elements is their propensity to
form a variety of stable 1D structures including linear
sulfur/selenium chains\cite{{DT223},{DT267}} and double-helices of
selenium~\cite{DT222}. These structures can be characterized by
bond angles $\gamma$ and dihedral angles $\psi$, which are defined
in Fig.~\ref{fig1}(e). $a$-helices, shown in Fig.~\ref{fig1}(a),
consist of Se atoms connected to two neighbors with a bond angle
$\gamma_{a}{\approx}103^{\circ}$ and a dihedral angle
${\psi_{a}}{\approx}100^{\circ}$, and can be found in bulk Se.
Through an untwisting process, an $a$-helix can be transformed to
a $c$-helix, as shown in Fig.~\ref{fig1}(c), where
$\gamma_{c}{\approx}113^{\circ}$ and
${\psi}_{c}{\approx}64^{\circ}$. By changing the dihedral angle,
the $a$-helix can be transformed to a $b$- or a
$d$-chain~\cite{DT267}, as shown in Fig.~\ref{fig1}(b) and (d). %
Two thirds of the bond angles in the $b$- and $d$-chains are found
at $\gamma{\approx}106^\circ$, the remaining one third at
$\gamma{\approx}103^\circ$. In the $b$-chain, one third of the
dihedral angles is found at $\psi{\approx}100^{\circ}$, and rest
at $\psi{\approx}90^{\circ}$, whereas in the $d$-chain, one third
of the dihedral angles is at $\psi{\approx}79^{\circ}$, and the
rest at $\psi{\approx}100^{\circ}$. Values close to the optimum
bond angle ${\gamma_{opt}}{\approx}106^\circ$ and the dihedral
angle ${\psi_{opt}}{\approx}83^\circ$, %
\modR{found earlier in unconstrained finite-length Se
chains~\cite{DT267},} %
explain the higher stability of the $b$ and $d$-chains in
comparison to the $a$- and $c$-helices by about 17~meV/atom.

\begin{figure}[b]
\includegraphics[width=1.0\columnwidth]{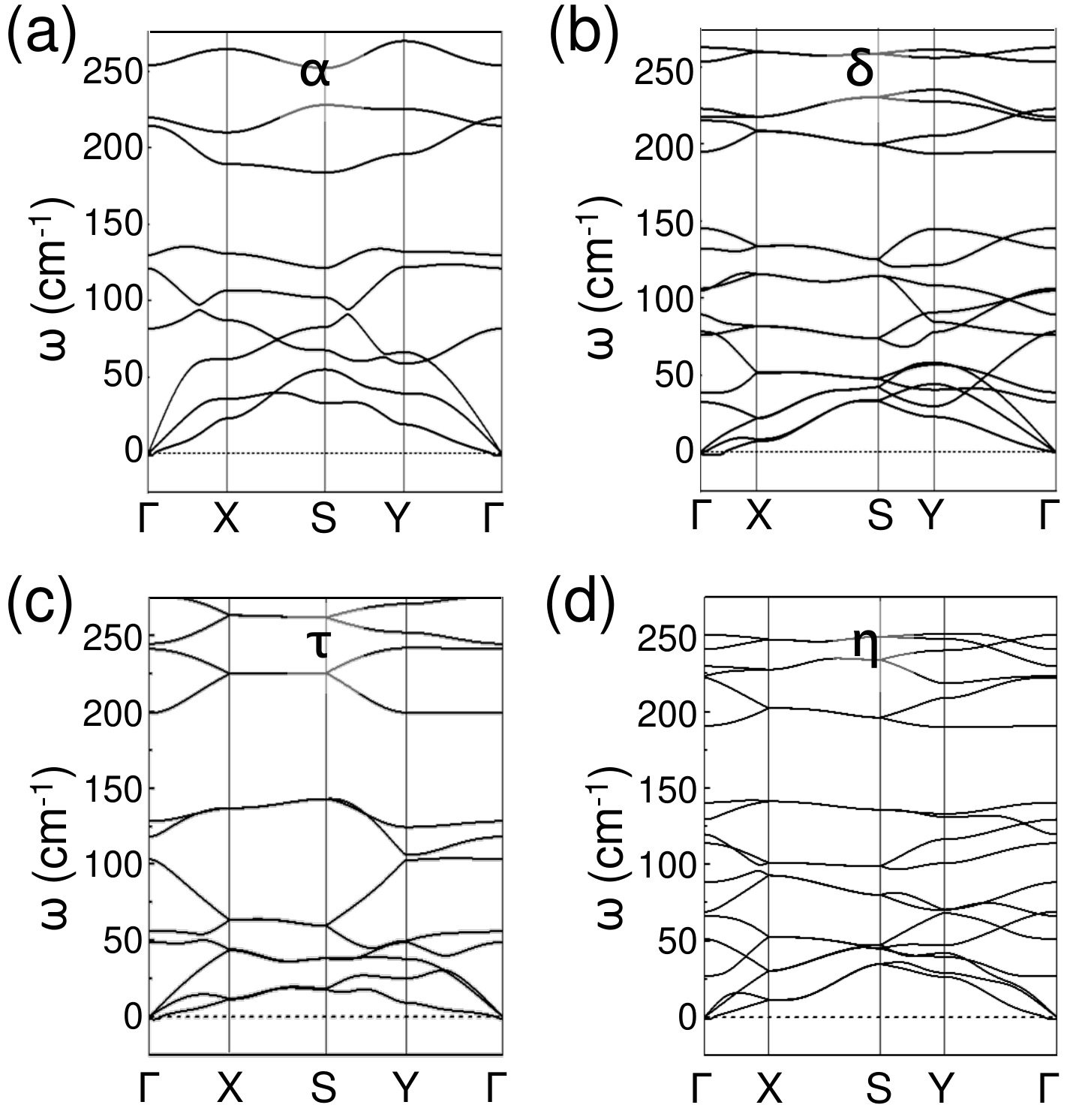}
\caption{%
\modR{ Phonon spectra of quasi-2D %
(a) $\alpha$-Se, %
(b) $\delta$-Se, %
(c) $\tau$-Se and %
(d) $\eta$-Se. } %
\label{fig2}}
\end{figure}

\begin{figure}[b]
\includegraphics[width=1.0\columnwidth]{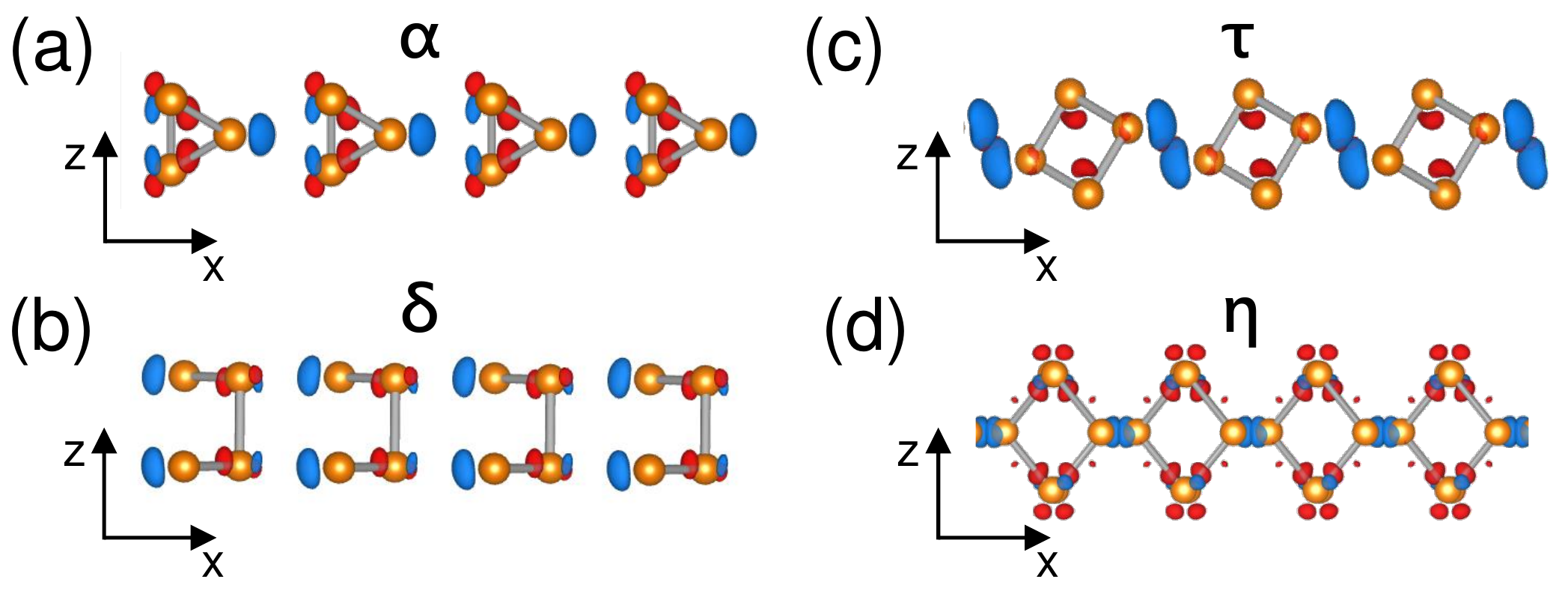}
\caption{%
Charge redistribution caused by connecting isolated 1D Se chains
to quasi-2D (a) $\alpha$-Se, (b) $\delta$-Se, (c) $\eta$-Se, and
(d) $\tau$-Se assemblies. Calling the optimum geometry of the
quasi-2D states the $A$ structure, we display the difference
charge density ${\Delta}\rho=\rho$({\rm A})$-\rho$({\rm
isolated-1D}) as isosurfaces bounding regions of electron excess
at $+2.5{\times}10^{-3}~\text{e}$/{\AA}$^3$, shown in red, and
electron deficiency at $-2.5{\times}10^{-3}~\text{e}$/{\AA}$^3$,
shown in blue. %
\label{fig3}}
\end{figure}

Facilitated by the multi-valence nature of Se~\cite{ZhangPRL2017},
these 1D Se structures can interact among each other, resulting in
the formation of quasi-2D allotropes. In parallel to the parent 1D
precursor $a$-helix, $b$-chain, $c$-helix and $d$-chain, we call
these quasi-2D allotropes $\alpha$-Se, $\delta$-Se~\cite{DT267},
$\tau$-Se, and $\eta$-Se~\cite{DT267}. Their unit cells are
rectangular, with the Bravais lattice vector $\vec{a}_1$ directed
along the chain axis and $\vec{a}_2$ in the inter-chain direction,
and are shown in Fig.~\ref{fig1}(f)-\ref{fig1}(i). %
\modR{The bulk Se structure may be viewed as an assembly of
$\alpha$-Se monolayers. Experimentally, only an $\alpha$-Se
bilayer has been synthesized successfully~\cite{Sarma2022}. }%
Numerical results for the cohesive energies and structural
parameters of the Se allotropes shown here are summarized in
Table~\ref{table1}. %

\modR{$E_{coh}$ values of all 1D and quasi-2D Se allotropes have
been calculated using total energy differences with respect to
isolated Se atoms.} %
\modR{The dynamical stability of the four quasi-2D Se allotropes
discussed in this work has been confirmed by the phonon spectra
displayed in Fig.~\ref{fig2}. As an independent proof of dynamic
stability, we have performed finite-temperature molecular dynamics
(MD) simulations of these systems. } %
\modB{ We have found the quasi-2D $\alpha$-Se, $\delta$-Se, and
$\eta$-Se structures to be stable at $T=300$~K, but $\tau$-Se to
be stable only at the lower temperature of $T=200$~K. Additional
information about the MD simulations is presented in Section D
of the Appendix. }%


The reason why we call these allotropes `quasi-2D' and not `2D' or
`1D' is that the inter-chain interaction within the layer is much
smaller than that of covalent bonds, but still stronger than the
inter-layer `vdW' interaction. Our DFT calculations based on the
PBE-D2 functional show that in comparison to isolated 1D
structures, the quasi-2D
allotropes gain %
${\Delta}E=117$~meV/atom in ${\alpha}$-Se, %
${\Delta}E=112$~meV/atom in ${\delta}$-Se, %
${\Delta}E=101$~meV/atom in ${\tau}$-Se, and %
${\Delta}E=125$~meV/atom in ${\eta}$-Se. %
This interaction also changes the bond angles $\gamma$ and the
dihedral angles $\psi$ in each 1D subsystem of these quasi-2D
allotropes. Typical changes are %
${\Delta}{\gamma}, %
{\Delta}{\psi}{\lesssim}2^{\circ}$ in $\alpha$-Se, %
${\Delta}{\gamma}{\lesssim}5^{\circ}$ and
${\Delta}{\psi}{\lesssim}8^{\circ}$ in $\delta$-Se, %
${\Delta}{\gamma}{\lesssim}5^{\circ}$ and
${\Delta}{\psi}{\lesssim}21^{\circ}$ in $\tau$-Se, and %
${\Delta}{\gamma}{\lesssim}2^{\circ}$ and
${\Delta}{\psi}{\lesssim}3^{\circ}$ in $\eta$-Se.

\modR{Assembly of isolated 1D structures to layers also leads to a
charge redistribution that may cause polarization changes in the
system. Bader charges of individual Se atoms, which were all the
same in isolated chains, change by %
${\Delta}Q{\approx}0.05$e in ${\alpha}$-Se,
${\Delta}Q{\approx}0.02$e in ${\tau}$-Se,
${\Delta}Q{\approx}0.04$e in ${\delta}$-Se, and
${\Delta}Q{\approx}0.05$e in ${\eta}$-Se. Such charge transfers
are a signature of forming local electric dipoles. Judging from
Fig.~\ref{fig3}(a)-\ref{fig3}(d), we can clearly see a depletion
of the electron density in the space in-between the chains and
charge redistribution among the inequivalent Se atoms. A similar
change in the charge density has been reported in bilayer
phosphorene~\cite{DT250}, where ${\approx}0.075$e per P atom have
been redistributed in comparison to a superposition to monolayer
charge densities, which would not be expected in a purely
vdW-bonded system. }%

\begin{figure*}[t]
\includegraphics[width=1.5\columnwidth]{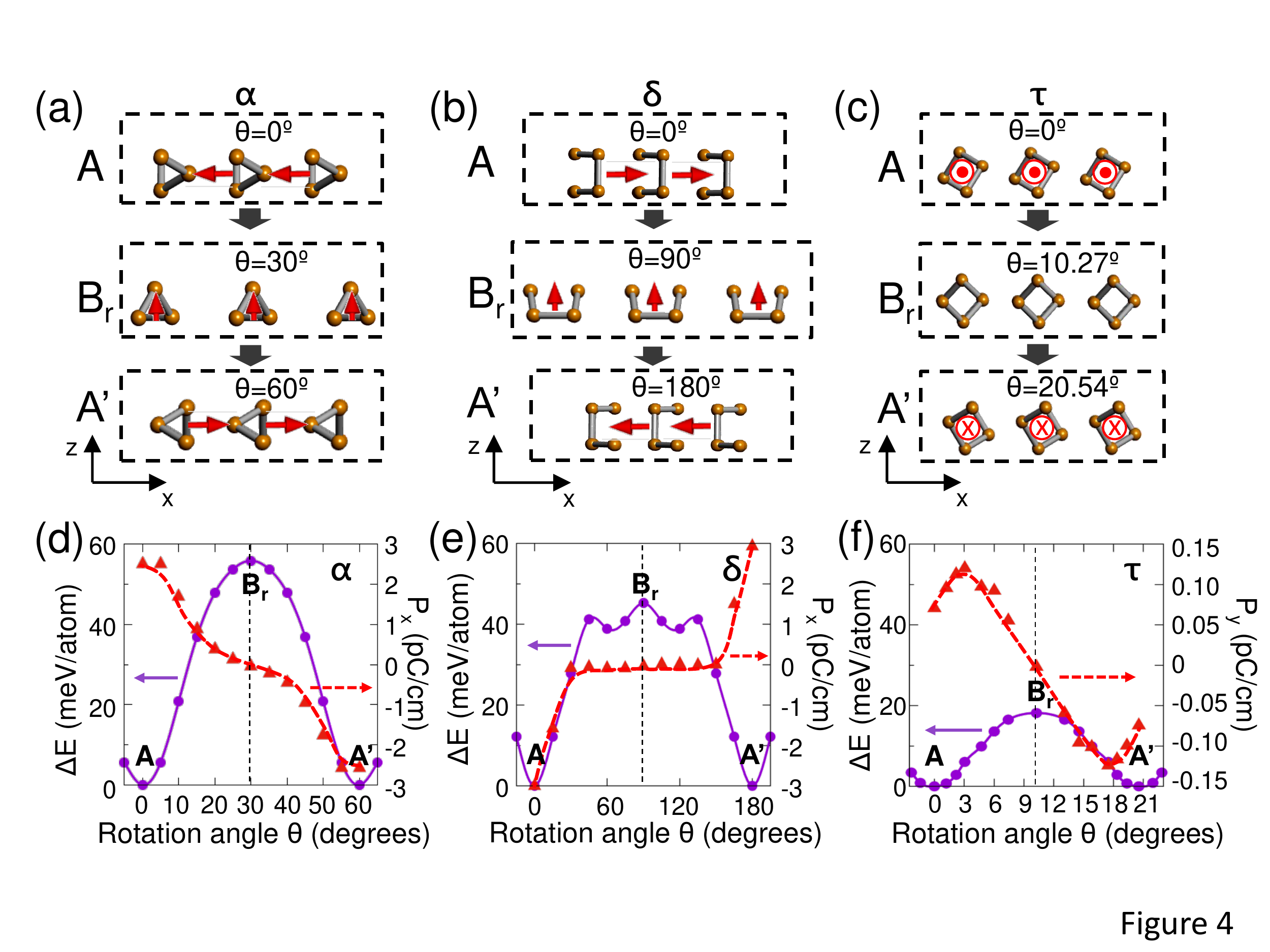}
\caption{%
Axial chain rotation process about the angle $\theta$ in quasi-2D
Se allotropes. Shown is the transition from the initial state $A$,
across the rotational barrier state $B_r$, to the closest
energetically degenerate state $A'$ with the opposite polarization
direction. (a)-(c) Morphology and polarization and (d)-(f) energy
changes in $\alpha$, $\delta$, and $\tau$ allotropes of Se. The
electric dipole moment is shown by the red arrow, with $\bigodot$
representing the out-of-plane and $\bigotimes$ the into-the-plane
direction. %
\label{fig4}}
\end{figure*}

\modR{Ferroelectricity in well-studied 3D systems such as
BaTiO$_{3}$~\cite{{Eom2004},{Duan2015}} and
PbTiO$_3$~\cite{{Noheda2011},{Vanderbilt2009}} is associated with
the %
\modB{ relative displacement of sublattices, which is
characterized by the position of a representative atom within the
unit cell}. In the most symmetric, but least stable geometry,
which we may call $B_S$ and which places this atom in the center
of the unit cell, makes the system a paraelectric (PE) with no
local dipole moments. Displacing this atom from the unstable $B_S$
state along a favorite direction to the most stable geometry $A$
results in an energy gain and formation of local electric dipoles
causing FE behavior. Displacing the atom in the opposite direction
to an equivalent geometry $A'$ only flips the direction of the
polarization by $180^\circ$. }%

\modR{The situation in our study is much more complex. We deal not
only with one atom within a unit cell, but with many atoms in a
chain segment that has an axial rotation degree of freedom. Also
the calculation of the polarization using the Berry phase approach
is more complex than in the above 3D systems and is presented in
detail in Appendix A. }%

\modR{Our calculations indicate that $\alpha$-Se and
${\delta}$-Se are ferroelectrics with the polarization values }%
$P=2.53$~pC/cm in $\alpha$-Se and %
$P=2.96$~pC/cm in ${\delta}$-Se. %
These values are comparable to values observed in typical 2D
monochalcogenide ferroelectrics of group IV
elements~\cite{{Nagashio2020},{Qian2017},{Guan2021}}, such as GeS,
SnS or SnSe. %
\modR{In $\tau$-Se, a net electric polarization of
$P_{\tau}$=0.07~pC/cm points along the $\vec{a}_1$ direction. In
the following, we will not discuss $\eta$-Se with an AFE behavior
and a vanishing net polarization, but will focus on $\alpha$-Se,
$\delta$-Se and $\tau$-Se with a non-vanishing polarization. }%

Within the chalcogen group, tellurium (Te) should be similar to Se
not only in its chemical behavior, but also in its electronic
properties. As could be expected, results published for $\beta$-Te
indicate that this 2D allotrope displays about half the
polarization found in Se, since systems containing this heavier
element are more metallic~\cite{Ji2018}.

\modR{Even though the inter-chain interaction in the 2D systems
discussed here is stronger than a pure vdW interaction, it is
weaker than a covalent interaction, thus allowing this rotational
degree of freedom to be activated at moderate energy cost. We may
thus expect gradual changes in the magnitude and direction of the
polarization when activating the axial rotation degree of freedom. }%

\subsection{Tunability of the polarization by chain rotation}

Unlike in uniform, traditional 2D materials, where all atoms are
connected by strong chemical bonds of typically
${\gtrsim}1$~eV/atom, the quasi-2D allotropes of Se we consider
here consist of 1D chains bonded by a weaker inter-chain
interaction of around $0.1$~eV/atom. This weaker interaction does
not restrict the rotation of individual constituent chains.
Whereas in-plane strain, %
\modB{ applied to a general 2D system, }%
changes the atomic structure, it usually maintains the symmetry.
Rotation of the individual chains by a nonzero angle $\theta$,
however, does modify the symmetry of the system. As a result, the
polarization of the system, which is related to symmetry, will
also be modified.

\begin{figure}[t]
\includegraphics[width=1.0\columnwidth]{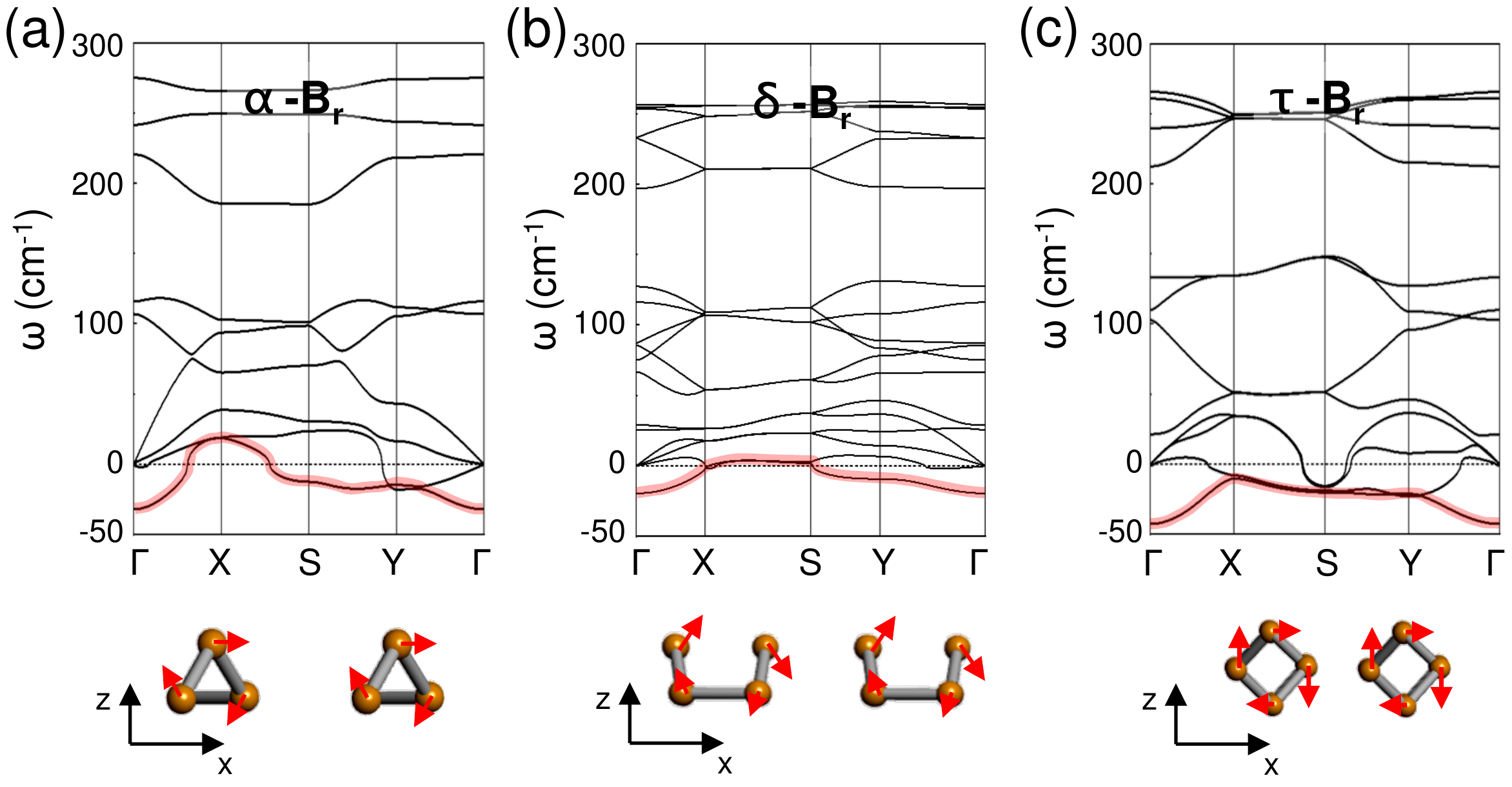}
\caption{%
\modR{Phonon spectra of the transition state structure $B_r$,
which is unstable with respect to the rotation of constituent
chains in (a) $\alpha$-Se, (b) $\delta$-Se and (c) $\tau$-Se. The
lowest branch of the phonon spectra with imaginary frequency is
highlighted by the red line. Corresponding modes, initiating
structural change, are indicated by the red arrows in the bottom
panels.}%
\label{fig5}}
\end{figure}

As also implied in Fig.~\ref{fig4}(a)-\ref{fig4}(c), the initial
ground state structure $A$ with $\theta=0^\circ$ is equivalent to
the $A'$ structure with the same symmetry for %
$\theta=60^{\circ}$ in $\alpha$-Se, %
$\theta=180^{\circ}$ in $\delta$-Se, and %
$\theta=20.54^{\circ}$ in $\tau$-Se. %
The energetically degenerate structures $A$ and $A'$ are related
to each other by a $180^{\circ}$ rotation of the system about an
axis normal to the plane. Thus, the polarization direction is
reversed between $A$ and $A'$.

Rotation of individual chains within the quasi-2D layer is an
energetically activated process, with the barriers related to the
inter-chain interaction. %
\modR{ For simpler systems, the transition path is usually
determined using the Nudged Elastic Band (NEB) approach.
Considering the number of atoms per unit cell, each with 3 degrees
of freedom, and the possibility of Bravais lattice vector changes,
optimizations would have to be performed in a roughly
20-dimensional configuration space, well beyond the scope of NEB.
To be tractable, we make specific assumptions about the path in
configuration space in an approach dubbed ``poor-man's NEB'' that
has been introduced previously~\cite{DT267}. For each rotation
angle, kept as a constraint, we optimized the inter-chain distance
and then the atomic positions. This approach may overshoot the
value of the activation barrier at $B_r$ between the between the
ground-state structures $A$ and $A'$, but is still useful to
characterize a likely path in configuration space involving chain
rotations that leads to changes in polarization. The values of the
activation barriers between equivalent configurations, determined
in this way, are 60~meV/atom for $\alpha$-Se, 50~meV/atom for
$\delta$-Se, and 20~meV/atom for $\tau$-Se, as shown in
Fig.~\ref{fig4}(d)-\ref{fig4}(f). We validated our approach by
calculating the phonon spectra at the unstable transition
structure $B_r$ and present our results in Fig.~\ref{fig5}. As
expected, we observe a soft mode with an imaginary frequency that
is characterized at $\Gamma$ by the rotation of the constituent 1D
chains towards the more stable $A$ or $A'$ structures in
$\alpha$-$B_r$, $\delta$-$B_r$ and $\tau$-$B_r$. }%

\modR{During the chain rotation from state $A$ to $A'$,
significant changes occur in the electronic structure and
especially the polarization. Our results indicate that both
out-of-plane and in-plane polarization components are changing,
with the polarization direction depicted in
Fig.~\ref{fig4}(a)-\ref{fig4}(c) and the value of the in-plane
component of the polarization shown in
Fig.~\ref{fig4}(d)-\ref{fig4}(f). } %

\modR{An independent Bader charge calculation indicates that
polarization changes are caused by charge reorganization during
the rotation of individual chains. We have found that rotation of
$a$-chains in $\alpha$-Se changes Bader charges on individual
atoms by ${\Delta}Q{\approx}0.01$e. According to
Fig.~\ref{fig4}(a), changing the rotation angle from geometry $A$
with ${\theta}=0^\circ$ to geometry $A'$ with ${\theta}=60^\circ$
flips the direction of the in-plane polarization. At the
transition state $B_r$ with ${\theta}=30^\circ$, the polarization
direction changes to out-of-plane with a value of $P=0.02$~pC/cm.
A similar change in polarization occurs in $\delta$-Se according
to Fig.~\ref{fig4}(b) during a rotation by
${\Delta\theta}=180^\circ$. For this system,
${\Delta}Q{\approx}0.03$e, and the transition state $B_r$ at
${\theta}=90^\circ$ acquires an out-of-plane polarization of
$P=0.017$~pC/cm. According to Fig.~\ref{fig4}(c), the behavior of
$\tau$-Se during the rotation of the constituent $c$-helices is
quite different. Unlike in the previous cases, the in-plane
polarization direction is along and not normal to the chains, but
still flips its direction during a rotation by
${\Delta\theta}=20.54^\circ$. At the activation barrier state
$B_r$ with ${\theta}=10.27^{\circ}$, the system turns into a
highly symmetric paraelectric with a vanishing polarization. }%

\begin{figure}[b]
\includegraphics[width=0.9\columnwidth]{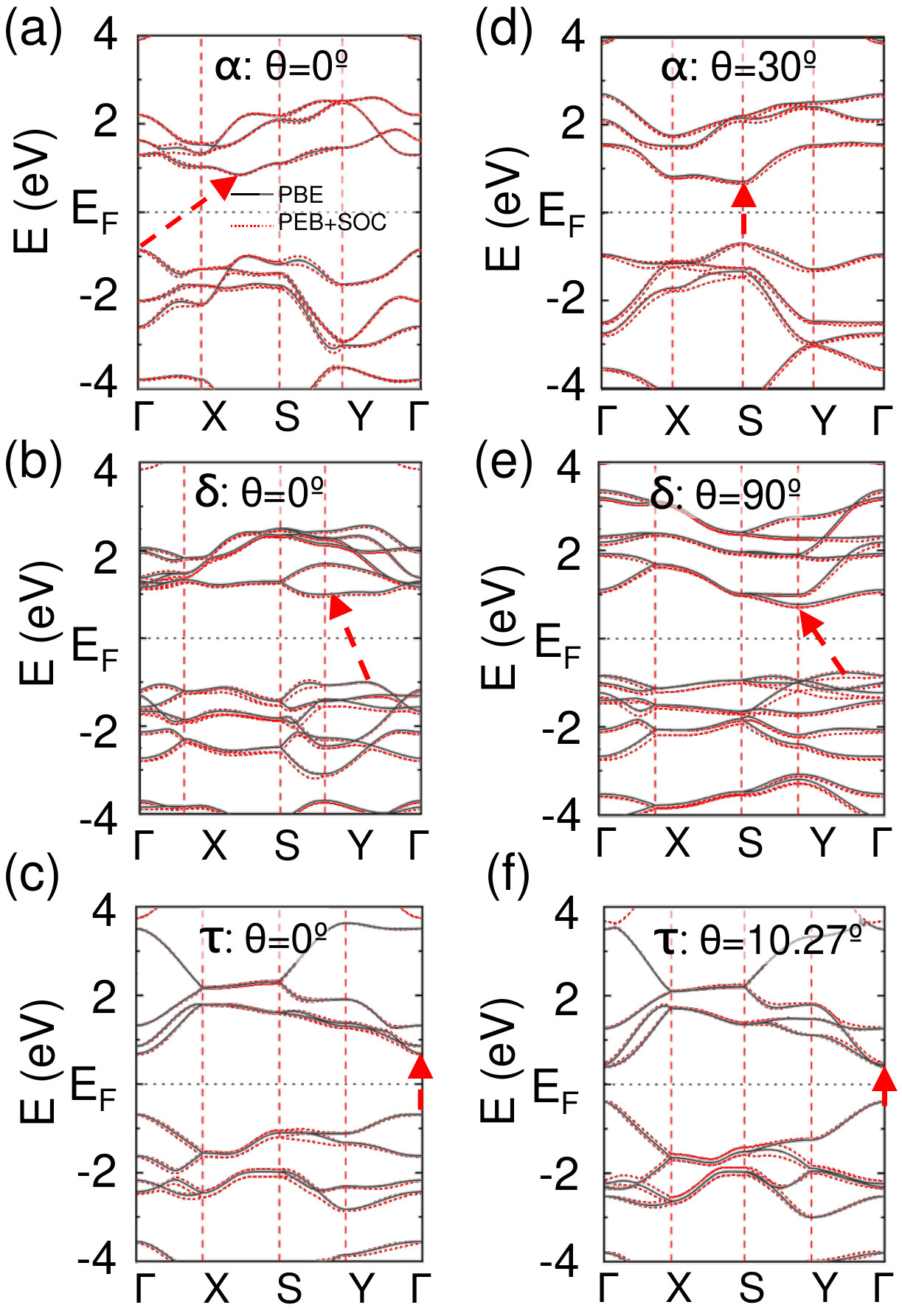}
\caption{%
Effect of the axial rotation angle $\theta$ of the constituent 1D
structures on the electronic band structure $E(\bf{k})$ in
quasi-2D
Se allotropes. $E(\bf{k})$ in %
$\alpha$-Se is shown for (a) $\theta=0^\circ$ and %
                         (d) $\theta=30^\circ$, %
in $\delta$-Se for (b) $\theta=0^\circ$ and %
                   (d) $\theta=90^\circ$, %
in $\tau$-Se   for (c) $\theta=0^\circ$ and %
                   (d) $\theta=10.27^\circ$. %
\modR{PBE-D2 results without spin-orbit coupling (SOC) are
presented by the solid black lines and those with SOC by the
dashed red lines. }%
\label{fig6}}
\end{figure}

\subsection{Tunability of the band structure by chain rotation}

Not only the net polarization, but also the band structure of the
system is affected by the chain rotation. Results of our DFT-PBE
calculations, presented in Fig.~\ref{fig6}(a)-\ref{fig6}(c),
suggest that all Se allotropes discussed here are semiconductors
with fundamental band gaps of $1.70$~eV in $\alpha$-Se, $1.99$~eV
in $\delta$-Se, and $1.38$~eV in $\tau$-Se. %
\modR{We have repeated these calculations by specifically
considering the effect of spin-orbit coupling (SOC). Since the
results with and without SOC, presented in Fig.~\ref{fig6}, are
very similar, we conclude that the role of SOC is not significant
in our systems. } %
Whereas the band gap
values are typically underestimated in DFT calculations, the band
dispersion is expected to be correct. Even though the systems are
strongly anisotropic, the band dispersion along the $\Gamma-X$
direction along the quasi-1D chains is not significantly smaller
than along the $\Gamma-Y$ direction normal to it. We expect
$\alpha$-Se and $\delta$-Se to be indirect-gap semiconductors,
whereas $\tau$-Se should have a direct band gap at the $\Gamma$
point. In $\alpha$-Se, rotating the individual $a$-helices by
$30^{\circ}$ reduces the fundamental band gap to $1.40$~eV and
turns it into a direct gap near the $S$-point, as seen in
Fig.~\ref{fig6}(d). As seen in Fig.~\ref{fig6}(e), a $90^{\circ}$
rotation of the $b$-chains in $\delta$-Se causes a reduction of
the band gap to $1.55$~eV, but keeps the gap indirect. According
to Fig.~\ref{fig6}(f), the band gap in $\tau$-Se remains indirect,
but is reduced to $0.80$~eV as the $c$-helices rotate by
$10.27^{\circ}$.

\begin{figure}[b]
\includegraphics[width=1.0\columnwidth]{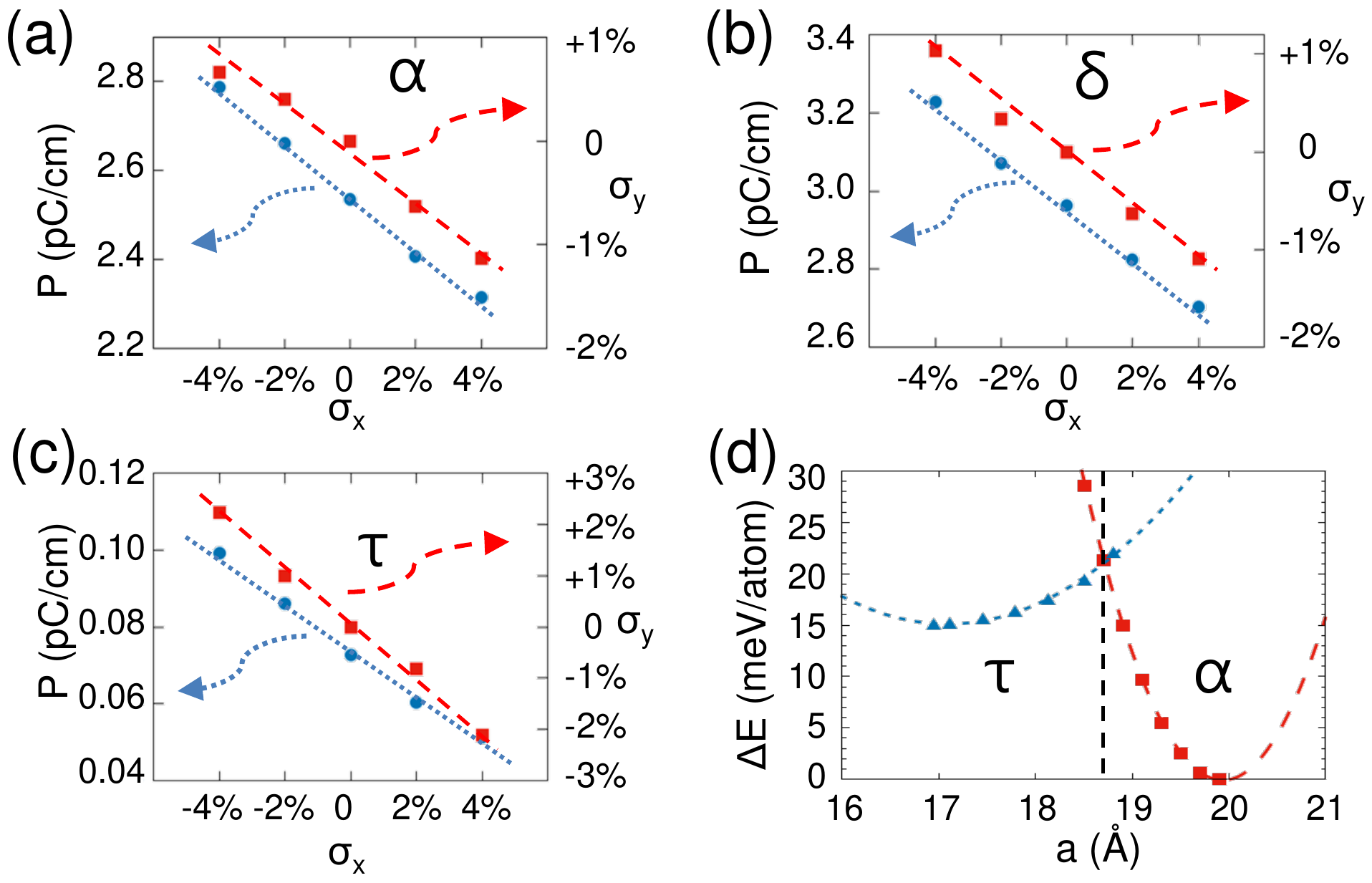}
\caption{%
Net polarization $P$ and deformation $\sigma_y$ along the
$\vec{a}_2$ direction as a function of uniaxial strain $\sigma_x$
along the $\vec{a}_1$ direction in quasi-2D (a) $\alpha$-Se (b)
$\delta$-Se, and (c) $\tau$-Se. The effect of $\sigma_x$ on $P$ is
shown by the blue dotted lines, and corresponding changes in
$\sigma_y$ are shown by the red dashed line. %
(d) Energy change ${\Delta}E$ caused by changing the lattice
constant $a$ in $\tau$-Se and $\alpha$-Se along the $\vec{a}_1$
direction. Results for supercells of $\alpha$-Se, with %
$a=4a_1{\rm({\alpha}-Se)}$, are shown by the red dashed line.
Results for supercells of $\tau$-Se, with %
$a=3a_1{\rm({\tau}-Se)}$, are shown by the blue dotted line. %
\label{fig7}}
\end{figure}

\subsection{Effect of in-plane strain on the polarization %
            of quasi-2D Se allotropes}

We also study the effect of in-plane uniaxial strain on the
polarization.
\modB{ To do so, we strain the lattice uniformly }%
along the $\vec{a}_1$ direction, which is aligned with the axes of
the constituent chains in $\alpha$-Se, $\delta$-Se and $\tau$-Se.
To reproduce the realistic %
\modB{ response of a system subject to this strain, }%
we allow the ${a}_2$ lattice constant to relax for each value of
${a}_1$. As the quasi-2D structures are stretched along the
$\vec{a}_1$ direction, they shrink linearly along the $\vec{a}_2$
direction, indicating a positive Poisson ratio.

\begin{figure*}[t]
\includegraphics[width=1.8\columnwidth]{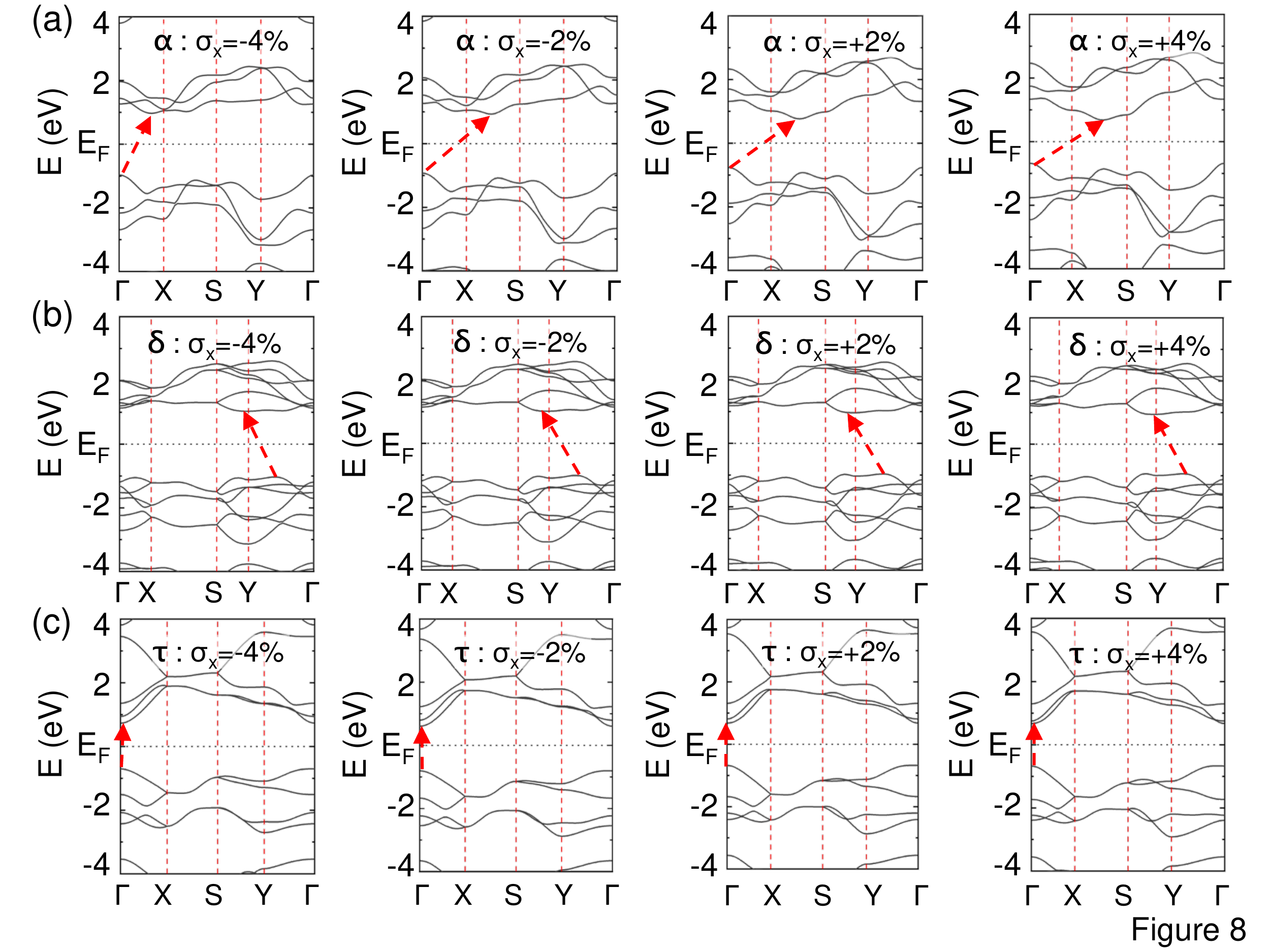}
\caption{%
Effect of the uniaxial strain $\sigma_x$ along the
chain direction on the electronic band structure $E(k)$ in (a)
$\alpha$-Se (b) $\delta$-Se, and (c) $\tau$-Se. Lowest-energy
transitions across the fundamental band gap are indicated by the
red dashed lines and arrows.%
\label{fig8}}
\end{figure*}

For $\alpha$-Se shown in Fig.~\ref{fig7}(a) and $\delta$-Se in
Fig.~\ref{fig7}(b), the change in the lattice constant ${a}_2$ is
about one fourth of the change in the lattice constant ${a}_1$
along the chain axis. The resulting Poisson ratios are
${\nu}{\approx}0.22$ for ${\alpha}$-Se and ${\nu}{\approx}0.27$
for ${\delta}$-Se. According to Fig.~\ref{fig7}(c), the Poisson
ratio of $\tau$-Se, ${\nu}{\approx}0.54$, is about twice that of
$\delta$-Se.

The net polarization as a function of strain is also shown in
Fig.~\ref{fig7}(a)-\ref{fig7}(c). These results indicate a linear
increase in the polarization as the structure is compressed along
the chain axis. For the uniaxial strain value $\sigma_x$=-4$\%$,
we observe an increase in the polarization by $9.8$\% in
$\alpha$-Se, $9.1$\% in $\delta$-Se, and by $37.5$\% in $\tau$-Se.

As can be inferred from Fig.~\ref{fig1}(a) and \ref{fig1}(c), the
$a$-helix component of $\alpha$-Se and the $c$-helix component of
$\tau$-Se are topologically related. They can be transformed from
one to another by a twisting/untwisting process, in which the bond
angles and the dihedral angles change. We should note that for
supercells containing the same number of atoms, the lattice
constant along the chain direction is $a=4a_1$ for ${\alpha}$-Se
and $a=3a_1$ for ${\tau}$-Se. The energy change ${\Delta}E$ as a
function of the supercell length in these two allotropes is shown
in Fig.~\ref{fig7}(d). We set the optimum structure of the more
stable $\alpha$-Se allotrope as the energy reference, with the
optimum $\tau$-Se structure being less stable by only
$15$~meV/atom. At the supercell lattice constant $a=18.72$~{\AA},
the energy of stretched $\tau$-Se equals that of compressed
$\alpha$-Se, indicating the possibility of a strain-induced phase
transformation at this point.

\subsection{Effect of in-plane strain on the band structure %
               of quasi-2D Se allotropes}

We also study the effect of in-plane uniaxial strain on the band
structure of $\alpha$-Se, $\delta$-Se and $\tau$-Se. Our results,
presented in Fig.~\ref{fig8}, indicate that the three allotropes
keep their semiconducting character when the quasi-1D structures
are axially compressed or stretched, with strain values $\sigma_x$
ranging from $-4\%$ to $+4\%$. Similar to our results for
unstained systems in Fig.~\ref{fig6}, $\alpha$-Se and $\delta$-Se
remain indirect-gap semiconductors, whereas $\tau$-Se has a direct
band gap at the $\Gamma$ point independent of strain. We find that
the fundamental band gap widens under compression and shrinks in
stretched systems. In particular, under $\sigma_x$=-4$\%$
compression,  $E_g$ increases by $14$\% in $\alpha$-Se, $3.5$\% in
$\delta$-Se and $3.6$\% in $\tau$-Se. Under $\sigma_x$=+4$\%$
stretch, $E_g$ decreases by $20$\% in $\alpha$-Se, $5.4$\% in
$\delta$-Se and $2.5$\% in $\tau$-Se. This behavior does not
follow the naive picture for 1D systems that stretching elongates
bonds, thus reducing hopping and the bandwidth, leading to wider
band gaps. In the quasi-2D systems we study, stretching the chains
reduces the inter-chain distance and thus increases the
inter-chain interaction. Since the band dispersion along and
normal to the chain direction is comparable, the naive picture of
1D systems no longer applies, resulting in dispersion changes seen
in Fig.~\ref{fig8}.

\section{Discussion}


We studied the possibility to tune the electric polarization of
four quasi-2D allotropes of Se. Besides $\alpha$-Se, the naturally
occurring phase consisting of $a$-helices, we also studied
$\delta$-Se, $\eta$-Se and $\tau$-Se, which differ only in their
dihedral angles and should be similarly stable. The transformation
of the $a$-helix to a $b$-chain, the constituent of $\delta$-Se,
has been discussed earlier~\cite{DT267}. A change of the dihedral
angle leads to the $d$-chain of $\eta$-Se, and untwisting the
$a$-helix leads to the $c$-helix of $\tau$-Se. Interaction of
these 1D structures should, as in the case of $\alpha$-Se, lead to
the formation of corresponding quasi-2D structures. %
\modR{ %
Since the
unit cells of these four allotropes are very different, selecting
a suitable substrate with a matching lattice constant should cause
preferential epitaxial growth of a given allotrope. }%
Charge redistribution due to the inter-chain interaction causes a
net electric polarization in many of these quasi-2D allotropes.

Our calculations of the polarization behavior in these unusual
systems are performed using a well-established
approach~\cite{Ji2018}, where we first identify a related, highly
symmetric paraelectric structure with no local dipoles, which we
call $B_S$. For the sake of completeness, we provide specific
details about the construction of the $B_S$ state in Appendix A.
The nature of the polarization behavior is then deduced by
comparing the charge distribution in a given system to that in the
$B_S$ state.

Changes of the electric polarization in ferroelectrics are caused
by relative displacement of sublattices. In most common
ferroelectrics including BaTiO$_3$, the interaction among atoms
forming such sublattices is indirect, effectuated by strong
covalent or ionic bonds to atomic neighbors in other sublattices.
In absence of a hierarchy of weaker and stronger interactions in
the system, a selective mechanical displacement of a sublattice
appears impossible.

The situation is very different in quasi-2D Se allotropes
discussed here, which consist of covalently bonded 1D constituents
that form sublattices. The weaker interaction between the 1D
systems provides one additional degree of freedom, namely the
axial rotation. This rotation, as we have shown, may change not
only the magnitude, but also the direction of the polarization
from in-plane to out-of-plane. We can imagine two scenarios to
effectuate such a rotation.

One of these scenarios uses nano-sized combs that have been
synthesized recently~\cite{{Yang2016},{Lee2014}}. We can imagine a
comb structure pressing from top onto a quasi-2D Se structure
adsorbed on a substrate. Inserting the `teeth' of the comb
in-between the chains and moving them along the chain direction
should result in a torque that would rotate each chain. The other
scenario is to place the quasi-2D system within a planar interface
in-between two solid blocks in a sandwich configuration. Applying
in-plane shear while pressing the blocks together should cause the
quasi-1D components to rotate like tree logs on a wagon. %
\modR{In both cases, we estimate that a torque
$>10^{-20}$~N$\cdot$m per unit cell should be required to initiate
the rotation of chains.} %
Strain transfer to the quasi-2D allotropes will be most efficient
when using solid blocks with a high Young's modulus such as
polydimethylsiloxane or polyvinyl
alcohol~\cite{{Liu2014},{Li2020}}.

We also imagine two scenarios to flip the direction of the
polarization in our quasi-2D Se allotropes. Both scenarios involve
atomic motion across a barrier in an energetically activated
process. In the first scenario, which involves a higher activation
energy and is discussed in more detail in Appendix C, atoms in a
rigid unit cell are displaced in opposite directions from the
initial state $A$, across a highly symmetric transition state
$B_S$, to the final state $A'$ with the same energy as $A$. This
process is shown in Fig.~\ref{fig11}(a)-\ref{fig11}(c). Since the
bond lengths and bond angles change significantly
during this process, the energy cost is rather high, amounting %
to $80.1$~meV/atom in ${\alpha}$-Se, %
$106.4$~meV/atom in ${\delta}$-Se, and %
$40.3$~meV/atom in ${\tau}$-Se %
according to Fig.~\ref{fig11}(d)-\ref{fig11}(f).

The second scenario involves a special rotation of the constituent
1D structures in the quasi-2D Se system, as illustrated in
Fig.~\ref{fig4}. In comparison to the first scenario, the energy
investment here is lowered to about %
$69.7\%$ for $\alpha$-Se, %
$42.5\%$ for $\delta$-Se, and %
$45.1\%$ for $\tau$-Se %
of its initial value. Thus, we may expect that applying an
external electric field may cause an axial rotation of rigid 1D
structures rather than atomic displacement within a rigid unit
cell. Clearly, lowering the inter-chain interaction should lower
the energy cost, allowing the chains to rotate more easily. On the
other hand, the inter-chain interaction must be significant enough
to cause charge redistribution and lead to polarization. We
believe that the quasi-2D Se allotropes discussed here bring a
good balance of both effects, providing the possibility to tune
the polarization by allowing for structural changes at a very
moderate energy cost.

The high flexibility of 1D Se chains in terms of bond length, bond
angle and dihedral angle suggests that many more structures beyond
the four allotropes discussed here may be realized, resulting in
other quasi-2D structures with very different electronic
properties.


\section{Summary and Conclusions}

In summary, we have performed {\em ab initio} DFT calculations to
study the polarization and band structure of four quasi-2D
allotropes of Se, which we called $\alpha$-Se, $\delta$-Se,
$\eta$-Se and $\tau$-Se. These allotropes are formed from their
one-dimensional (1D) constituents, which we call the $a$-helix,
$b$-chain, $c$-helix and $d$-chain, and which possess a rotational
degree of freedom about their axis. The inter-chain interaction is
not strong enough to suppress the rotation of these chains, but
still causes a charge redistribution within the quasi-2D
structures that results in a net polarization. $\alpha$-Se and
$\delta$-Se display an FE behavior with an in-plane polarization,
directed normal to the 1D chains. An in-plane polarization also
occurs in $\eta$-Se, which displays an AFE behavior. $\tau$-Se
behaves as a non-collinear ferrielectric, with FE polarization
along the chain axis and AFE polarization in the out-of-plane
direction. Rotating each constituent 1D chain in these quasi-2D Se
allotropes changes the atomic symmetry and also the electronic
structure. In the $\alpha$ and $\delta$ allotropes, chain rotation
changes the polarization direction from in-plane to out-of-plane.
At specific rotation angles, $\tau$-Se distorts to a highly
symmetric structure and its FiE polarization disappears. Along
with changes in polarization, caused by chain rotation, come
changes in the electronic band structure of these semiconducting
allotropes including a modification of the fundamental band gap,
which may change from an indirect to a direct gap in $\delta$-Se.
In-plane strain along the axial direction of the 1D structures
only modifies
the magnitude of the polarization.

\section*{Appendix}

\renewcommand\thesubsection{\Alph{subsection}}
\renewcommand{\theequation}{A\arabic{equation}}
\setcounter{subsection}{0} %
\setcounter{equation}{0} %
\setcounter{video}{0} %

%

\subsection{Polarization calculation using the Berry phase approach}

\begin{figure*}[t]
\includegraphics[width=1.9\columnwidth]{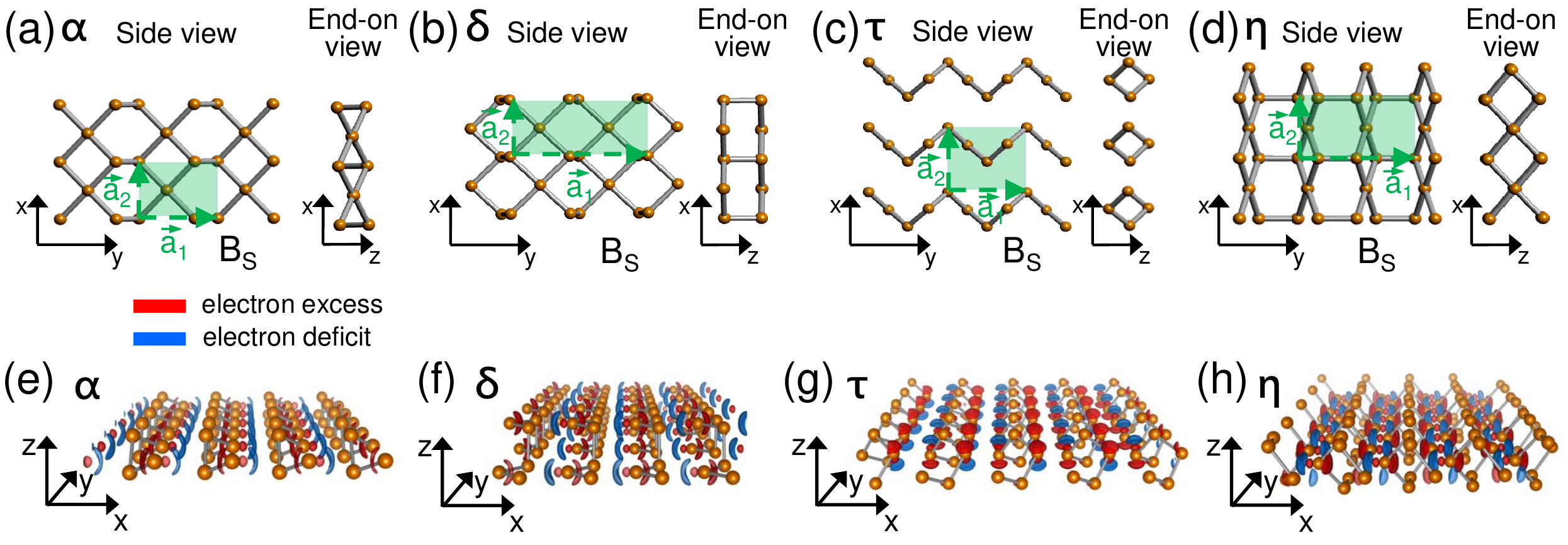}
\caption{Highly symmetric paraelectric phase $B_S$ of quasi-2D
allotropes of Se placed in the $x-y$ plane. The structure of %
(a) $\alpha$-Se, %
(b) $\delta$-Se, %
(c) $\tau$-Se, and %
(d) $\eta$-Se %
is shown in the side and the end-on view. The unit cells are
highlighted by the transparent green areas. %
Panels (e)-(h) display the charge redistribution
${\Delta}\rho=\rho$($A$)$-\rho$($B_S$) between the polarized phase
of %
$\alpha$-Se, %
$\delta$-Se, %
$\tau$-Se, and %
$\eta$-Se %
and their highly symmetric PE phase with no local dipoles, which
we call the $B_S$ state in the text.
Isosurfaces of ${\Delta}\rho$ are represented by bounding regions
of electron excess at $+7{\times}10^{-2}~\text{e}$/{\AA}$^3$ (red)
and electron deficiency at $-7{\times}10^{-2}~\text{e}$/{\AA}$^3$
(blue). The electric polarization is thus pointed from the red to
the blue region. %
\label{fig9}}
\end{figure*}

\modR{Polarization studies of a given allotrope using the Berry
phase approach begin with the construction of a highly symmetric
PE counterpart of this allotrope with no local electric dipoles.
Specifics of the construction process are outlined in Appendix B.
For the allotropes in this study, we display the morphology of the
PE state, characterized by the symbol $B_S$, in Fig.~\ref{fig9},
and determine their charge distribution. Structural optimization
of the $B_S$ state in these allotropes
causes a a reduction of the space group symmetry %
from $P2m$ to $P2$ in ${\alpha}$-Se, %
from $Pmma$ to $Pma2$ in ${\delta}$-Se, and %
from $Pmma$ to $P2_{1}$ in ${\eta}$-Se. %
This essentially translates to loss of mirror symmetry with
respect to a plane normal to the inter-chain direction. }%

\modR{Comparing the charge difference between these quasi-2D
allotropes and their corresponding $B_S$ states, we observe an
parallel electrical polarization along the $\vec{a}_2$ direction
of ${\alpha}$-Se and ${\delta}$-Se indicating FE behavior, shown
in Fig.~\ref{fig9}(e) and (f). And an antiparallel electrical
polarization along the $\vec{a}_2$ direction of ${\eta}$-Se
indicating an AFE behavior, shown in Fig.~\ref{fig9}(h). A special
behavior occurs in the unit cell of $\tau$-Se. In comparison to
the $B_S$ structure, two positively charged atoms both move in the
same direction along $\vec{a}_1$, but in opposite directions
normal to the plane.}
The space group symmetry %
$P222_{1}$ reduces to $P2_{1}$ in ${\tau}$-Se, %
as the rotation symmetry about $\vec{a}_2$ and about the direction
normal to the layer in the $B_S$ structure is lost, with only the
rotation symmetry about $\vec{a}_1$ remaining. %
\modR{As a result, $\tau$-Se displays a non-collinear polarization
behavior, indicated in Fig.~\ref{fig9}(g). We use the term `FiE',
which is defined for the similar non-collinear polarization in a
$\delta$-GeS monolayer~\cite{Guan2021}, to describe the
polarization behavior in $\tau$-Se. The polarization of the
quasi-2D Se allotropes discussed here is shown by the red arrows
in Fig.~\ref{fig1}(f)-\ref{fig1}(i). }%

\subsection{Construction and characterization of the PE state %
            in quasi-2D Se allotropes}

\begin{figure}
\includegraphics[width=1.0\columnwidth]{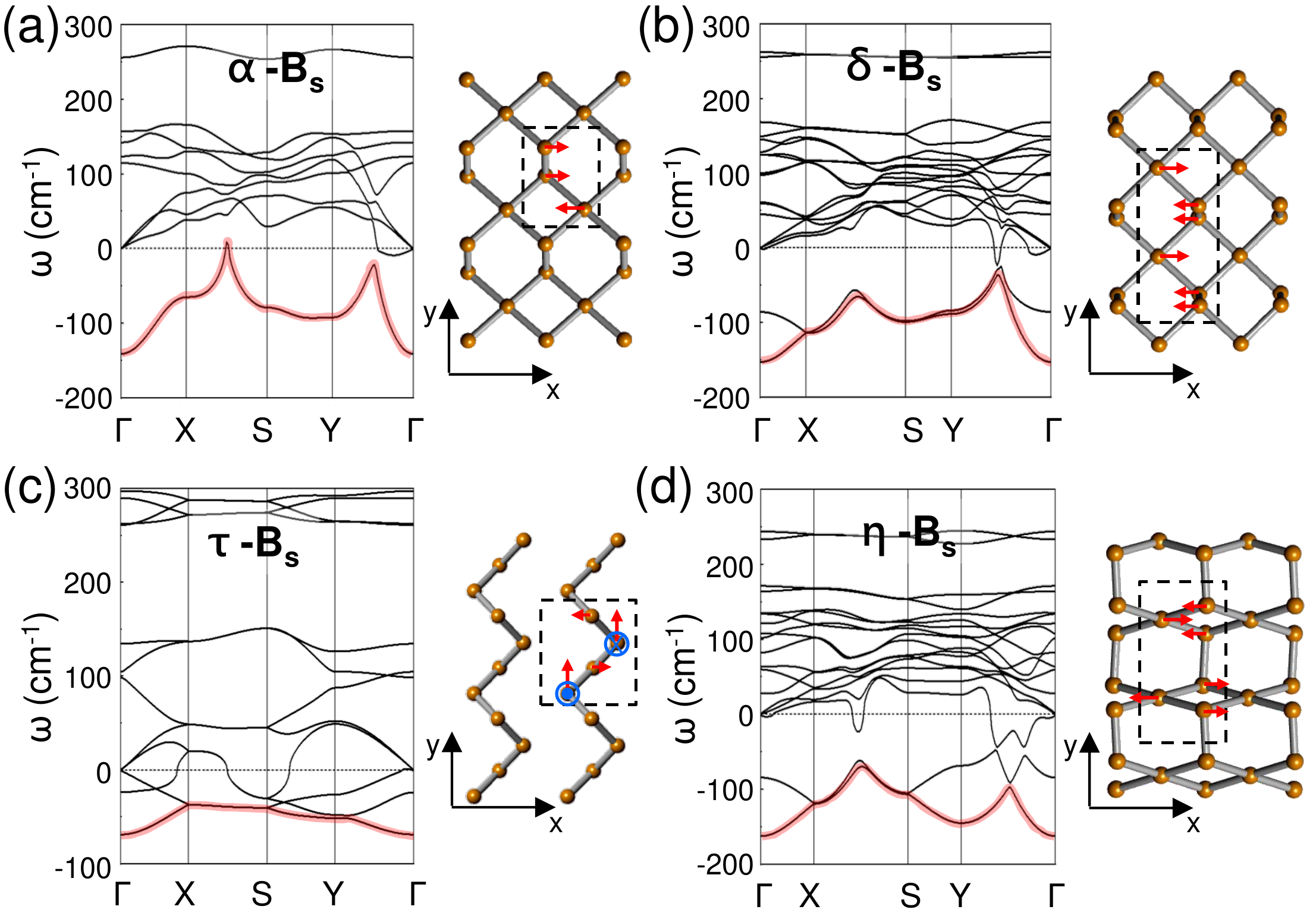}
\caption{ %
\modR{ Phonon spectra of the paraelectric phase $B_s$ of (a)
$\alpha$-Se, (b) $\delta$-Se, (c) $\tau$-Se and (d) $\eta$-Se. The
vibration mode of the lowest branch of the phonon spectra which is
highlighted by red line is represent by the red arrow in the right
panel. $\bigodot$ and $\bigotimes$ in (c) representing the
out-of-plane and the into-the-plane displacement
of Se atoms. }%
\label{fig10}}
\end{figure}

\begin{figure}
\includegraphics[width=1.0\columnwidth]{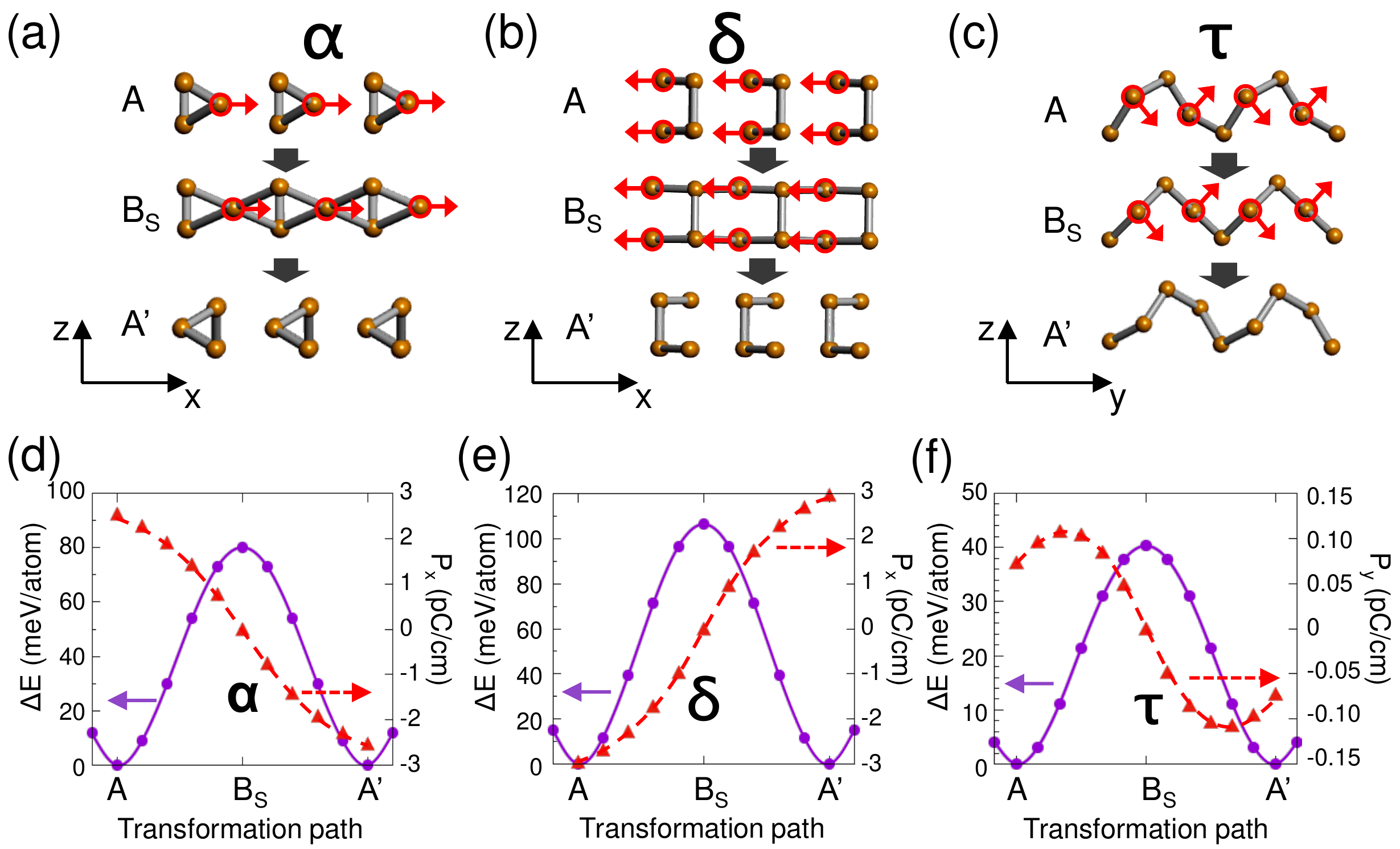}
\caption{Structural changes causing a flip of the polarization
direction in quasi-2D allotropes of Se. Atomic displacement,
indicated by the red arrows, starts from from the initial state
$A$, crosses the paraelectric state $B_S$, and is completed in the
final state $A'$ of (a) $\alpha$-Se, (b) $\delta$-Se
and (c) $\tau$-Se. %
Corresponding energy changes ${\Delta}E$ in (d) $\alpha$-Se,
(e) $\delta$-Se, and (f) $\tau$-Se. %
\label{fig11}}
\end{figure}

We use a traditional approach to construct the highly symmetric
paraelectric state called $B_S$. We start from the initial
structure $A$ and identify the polarization direction. We reflect
$A$ about a mirror plane that is perpendicular to the polarization
direction to arrive at $A'$. Since $A$ and $A'$ are mirror
symmetric, the polarization of $A'$ is opposite to that of $A$. A
state mid-way between $A$ and $A'$ has zero polarization and is
identified as the paraelectric reference state $B_S$. As can be
seen in Fig.~\ref{fig9}, such paraelectric structures have a
higher symmetry in comparison to the initial state $A$ shown in
Fig.~\ref{fig1}. %

\modR{We calculate the phonon spectra of these paraelectric
structures $B_S$ and show the results in Fig.~\ref{fig10}. The
phonon mode associated with the largest imaginary frequency at the
$\Gamma$-point indicates a decay to more stable structures $A$
and $A'$.} %

\subsection{Flipping the polarization by atom displacement}

One possible way to flip the polarization direction in the
quasi-2D allotropes by $180^{\circ}$ is to displace specific atoms
within the unit cell in a particular way. Let us focus on
$\alpha$-Se, $\delta$-Se and $\tau$-Se with a nonzero
polarization. Starting from the initial state $A$, we displace
specific atoms along the direction indicated by the red arrows in
Fig.~\ref{fig11}. The displacement continues across the barrier
state $B_S$ shown in Fig.~\ref{fig9} and completes at the state
$A'$. The stability of $A$ and $A'$ is the same, but the
polarization is opposite. The atomic displacement process causing
a flip of the polarization direction is shown in %
Fig.~\ref{fig11}(a) for $\alpha$-Se, %
Fig.~\ref{fig11}(b) for $\delta$-Se, and %
Fig.~\ref{fig11}(c) for $\tau$-Se. %
The energy barrier in this process is %
${\Delta}E{\approx}80.08$~meV/atom for ${\alpha}$-Se, %
${\Delta}E{\approx}106.41$~meV/atom for ${\delta}$-Se, and %
${\Delta}E{\approx}40.3$~meV/atom for ${\tau}$-Se, %
as seen in Fig.~\ref{fig11}(d)-\ref{fig11}(f). %
\modR{The gradual change of polarization caused by atom
displacement is shown by the red dashed lines in
Fig.~\ref{fig11}(d)-\ref{fig11}(f). Starting from the most stable
state $A$, the polarization first decreases to zero at the
paraelectric reference state $B_s$, and then increases to the
initial value with opposite direction at the final state $A'$.}

\subsection{Finite-temperature MD simulations of
            quasi-2D Se structures}

\begin{video}[h]
\includegraphics[width=0.6\columnwidth]{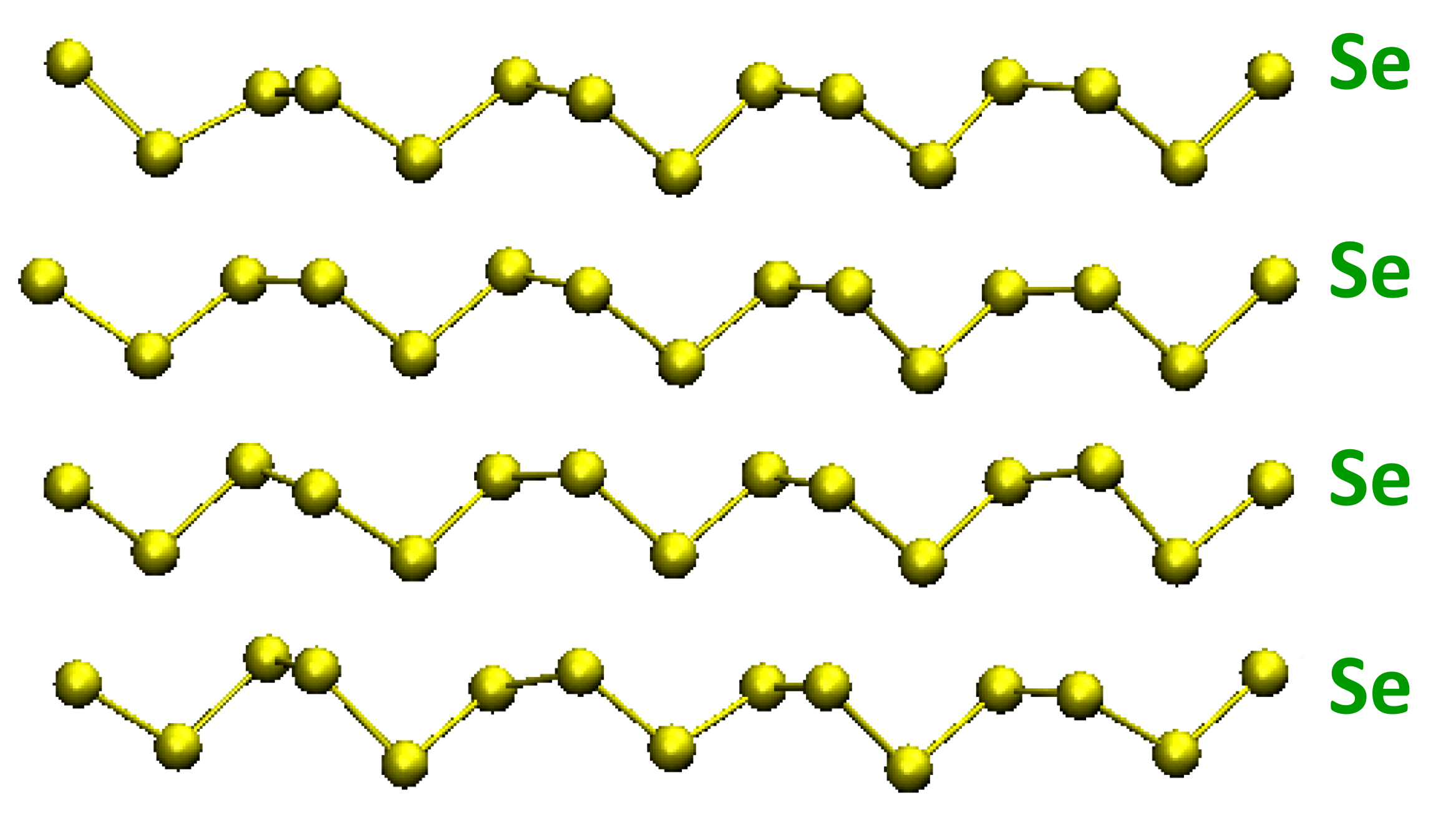}
\setfloatlink{video1.mp4} %
\caption{\modR{Results of a $2$~ps long MD simulation of
$\alpha$-Se at $T=300$~K.}%
\label{Video1}}
\end{video}

\modR{ We have performed canonical MD simulations at finite
temperatures as an independent proof of the thermodynamical
stability of the quasi-2D $\alpha$-Se, $\eta$-Se and $\tau$-Se
structures described in the main text. Results of $2$~ps long runs
at $T=300$~K are shown in Video~\ref{Video1} for $\alpha$-Se and
Video~\ref{Video2} for $\eta$-Se. Results of a $2$~ps run for
$\delta$-Se at $T=300$~K were published earlier~\cite{DT267}. As
can be inferred from Figs.~\ref{fig4} and \ref{fig11}, the energy
barrier in $\tau$-Se is much smaller then in $\alpha$-Se and
$\delta$-Se. We have also found $\tau$-Se to become unstable at
$300$~K. Therefore, we present results of a $2$~ps run for
$\tau$-Se in Video~\ref{Video3} at a lower temperature $T=200$~K.}

\begin{video}[t]
\includegraphics[width=0.6\columnwidth]{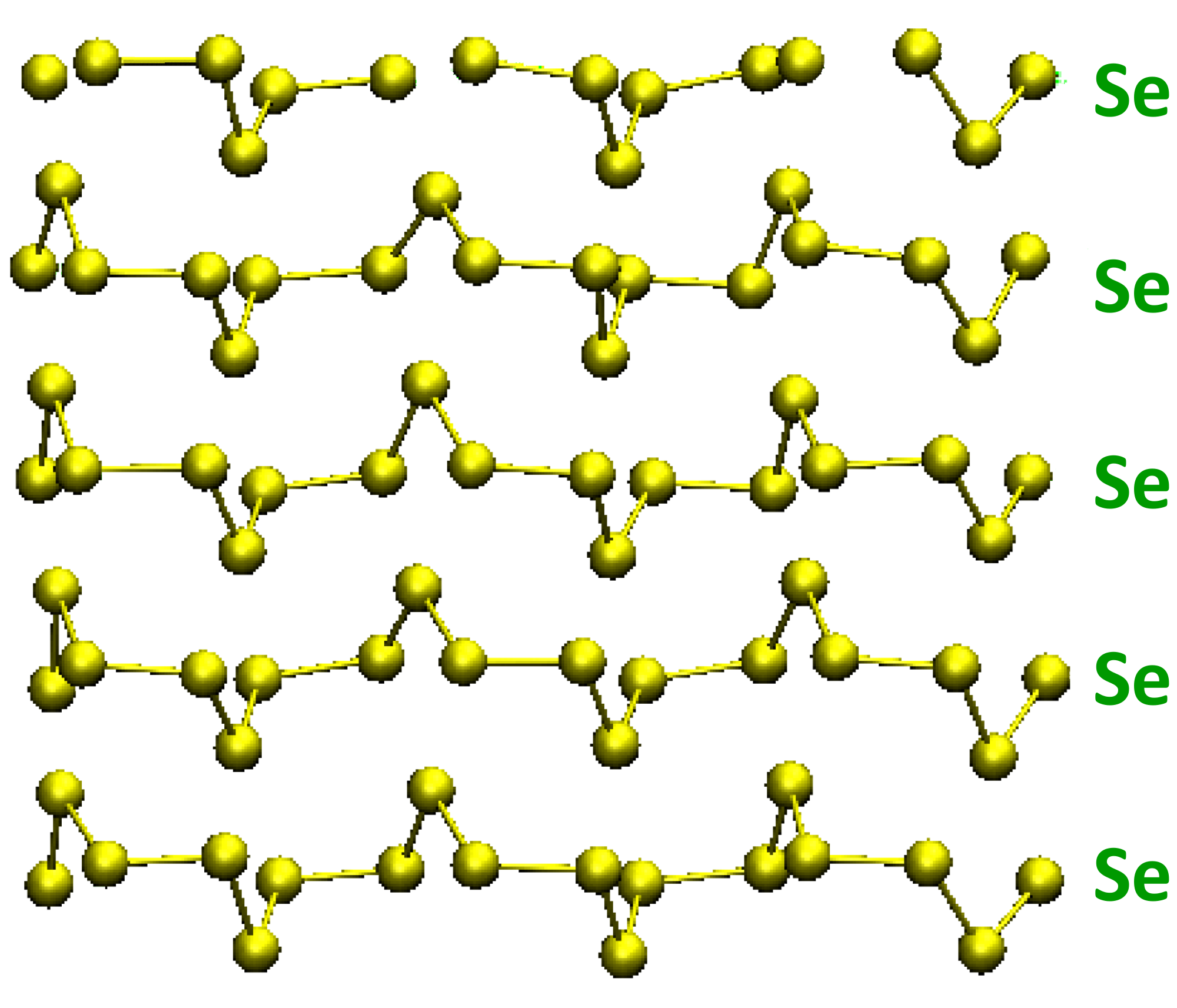}
\setfloatlink{video2.mp4} %
\caption{\modR{Results of a $2$~ps long MD simulation of $\eta$-Se
at $T=300$~K.}%
\label{Video2}}
\end{video}
\begin{video}
\includegraphics[width=0.6\columnwidth]{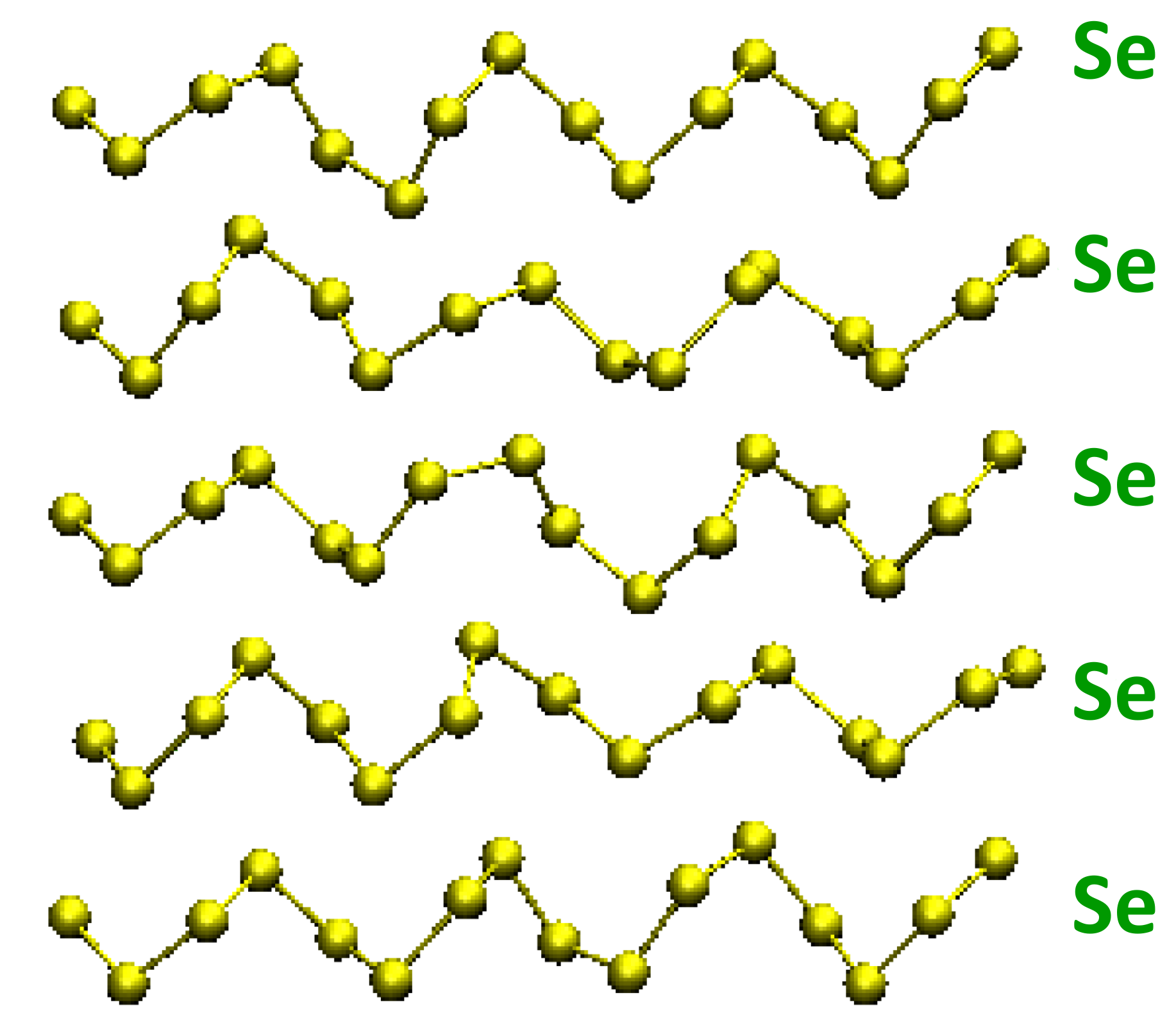}
\setfloatlink{video3.mp4} %
\caption{\modR{Results of a $2$~ps long MD simulation of $\tau$-Se
at $T=200$~K.}%
\label{Video3}}
\end{video}


\begin{acknowledgements}
We appreciate valuable discussions with Bei Zhao and Wei Liu about
the rotation strategy. D. Liu acknowledges financial support by
the Natural Science Foundation of the Jiangsu Province Grant No.\
BK20210198. R. Wei and S. Dong acknowledge financial support by
the National Natural Science Foundation of China (NNSFC) Grant
No.\ 11834002. S. Song and J. Guan acknowledge financial support
by NNSFC Grant No.\ 61704110. Computational resources for most
calculations have been provided by the Michigan State University
High Performance Computing Center.
\end{acknowledgements}


%
%
%


%

\end{document}